\documentclass[english]{scrartcl}

\usepackage{fullpage}
\usepackage{hyperref}  
\usepackage{url}
\usepackage{amsfonts,amsmath}
\usepackage{graphicx}
\usepackage{multirow}

\title{On numerical approximation schemes\\ for expectation propagation}

\date{}
\author{Alexis Roche\thanks{alexis.roche@\{centraliens.net,gmail.com\}}}

\def\btheta{{\boldsymbol{\theta}}}
\def\bmu{{\boldsymbol{\mu}}}
\def\bphi{{\boldsymbol{\phi}}}
\def\balpha{{\boldsymbol{\alpha}}}
\def\S{{\boldsymbol{\Sigma}}}
\def\e{{\mathbf{e}}}
\def\x{{\mathbf{x}}}

\begin{document}

\maketitle

\begin{abstract}
Several numerical approximation strategies for the expectation-propagation algorithm are studied in the context of large-scale learning: the Laplace method, a faster variant of it, Gaussian quadrature, and a deterministic version of variational sampling ({\em i.e.}, combining quadrature with variational approximation). Experiments in training linear binary classifiers show that the expectation-propagation algorithm converges best using variational sampling, while it also converges well using Laplace-style methods with smooth factors but tends to be unstable with non-differentiable ones. Gaussian quadrature yields unstable behavior or convergence to a sub-optimal solution in most experiments.
\end{abstract}

\section{Introduction}
\label{sec:introduction}

Many machine learning tasks involve minimizing an objective that has the form of a sum:
\begin{equation}
\label{eq:additive_objective}
u(\btheta) = \sum_{k=1}^n u_k(\btheta),
\end{equation}
where $\btheta\in\mathbb{R}^d$ is an unknown vector of interest, and $u_k(\btheta)$, for $k=1,\ldots,n$, are cost functions most often associated with distinct data points, or subsets of points called ``mini-batches''. Problems of type~(\ref{eq:additive_objective}) include maximum likelihood parameter estimation under independent measurements, empirical risk minimization \cite{Vapnik-00}, or composite likelihood maximization \cite{Varin-11}.  

While generic optimization algorithms may scale poorly with data size, it is possible to devise fast optimization procedures by taking advantage of the additive nature of~(\ref{eq:additive_objective}). For instance, stochastic gradient descent (SGD) methods \cite{Bottou-10} have a natural implementation using a single cost gradient~$\nabla u_k(\btheta)$ at each iteration rather than the total gradient~$\nabla u(\btheta)$. This makes the computational cost of an iteration proportional to the associated mini-batch size rather than to the full data size, hence providing a massive gain in efficiency over deterministic gradient methods in large-scale learning situations. 

SGD, however, assumes continuously differentiable cost functions. There are problems of interest where this assumption does not hold, such as support vector machine training \cite{Ertekin-11} or intensity-based image registration \cite{ijist:00}. In such context, a potential alternative to~SGD is the expectation-propagation (EP) algorithm \cite{Minka-01,Minka-05}, which is also scalable to big data while relying on larger-scale function approximations. EP is fundamentally an algorithm to approximate a factorial probability distribution, providing, in particular, an approximation to the mode. Hence, EP can be used to approximately minimize an objective of the form~(\ref{eq:additive_objective}) by converting it into a Boltzmann distribution:
\begin{equation}
\label{eq:boltzmann_dist}
p(\btheta) 
\equiv \frac{1}{Z(\beta)} e^{-\beta u(\btheta)}
\propto  \prod_{k=1}^n f_k(\btheta),
\qquad {\rm with} \quad
f_k(\btheta) \equiv e^{-\beta u_k(\btheta)}
,
\end{equation}
for some sufficiently large constant $\beta>0$. As in simulated annealing \cite{Kirkpatrick-84}, $\beta$ could be progressively increased for exact minimization, however it will be assumed fixed in this paper.

EP exploits the decomposition of~$p(\btheta)$ as a product of factors by alternatively fitting each factor~$f_k(\btheta)$ with an unnormalized Gaussian~$g_k(\btheta)$ via local moment matching. To approximate factor~$f_k$, EP~first forms the ``cavity distribution'',
\begin{equation}
\label{eq:cavity}
c_k(\btheta) = \prod_{\ell\not=k} g_\ell(\btheta),
\end{equation} 
which is the current approximation to the product of all factors except the one under consideration. The factor approximation~$g_k$ is then computed so that~$c_k g_k$ has the same moments of order~0, 1 and 2 as~$c_k f_k$. We will restrict ourselves to fully factorized Gaussian approximations, in which case cross second-order moments are ignored \cite{Minka-05}, leading to an~EP variant referred to as fully factorial~EP (FF-EP) in the sequel.

Since the moments essentially depend on the factor shape in the neighborhood of the cavity, the successive factor approximations are performed at adaptive, non-infinitesimal scales induced by the changing cavity distributions. Importantly, and similarly to~SGD, the complexity of an EP~iteration is proportional to the mini-batch size since a single factor is visited at a time. EP~variants such as Averaged~EP \cite{Dehaene-16} and Stochastic~EP \cite{Li-15} can further save memory load by constraining all factor approximations to be identical, but rest upon the same moment matching scheme. 

In comparison with~SGD, the problem of computing the gradient of~$u_k$ is replaced with that of computing the vector-valued integral of~$c_k f_k$, which is generally well defined but may lack a closed-form expression. Unless this can be worked around by applying a functional transformation to the factor, as in the Power~EP algorithm \cite{Minka-04b,Minka-05}, numerical approximations are needed. Both the Laplace method \cite{Smola-03,Yu-06} and Gaussian quadrature \cite{Zoeter-05,Yu-06} are common choices. I review these methods and introduce another method based on variational sampling \cite{ijasp:13}, which combines Gaussian quadrature with variational approximation, and is found empirically to yield better convergence of FF-EP with non-continuously differentiable factors.

\section{Numerical approximation schemes for~FF-EP}
\label{sec:approx_schemes}

The basic EP~building block consists of fitting an unnormalized Gaussian distribution to a single factor. Let us assume that a given factor~$f_k(\btheta)$ is selected, and drop the index~$k$ for clarity. The associated cavity distribution~$c(\btheta)$ is computed from the other factor current approximations according to~(\ref{eq:cavity}), and is a product of one-dimensional Gaussians by construction; let $\bmu$ and $\S = {\rm diag}(\sigma^2_1,\ldots,\sigma^2_d)$ denote its mean and variance matrix, respectively. 

In FF-EP \cite{Minka-05}, the $(2d+1)$ moments of~$c(\btheta)f(\btheta)$ up to the order~2 need to be computed, 
\begin{equation}
\label{eq:moments}
m_i = \int c(\btheta) f(\btheta) \phi_i(\btheta)d\btheta,
\qquad i=0,\ldots, 2d,
\end{equation}
for the monomials of degree $0, 1, 2$:
\begin{equation}
\label{eq:monomials}
\left\{
\begin{array}{ll}
\displaystyle \phi_0(\btheta) = 1, & \\
\displaystyle \phi_i(\btheta) = \theta_i, & i=1,\ldots,d \\
\displaystyle \phi_i(\btheta) = \theta^2_{i-d}, & i=d+1,\ldots,2d.
\end{array}
\right
.
\end{equation}

The factor approximation~$g(\btheta)$ is then readily given by $g(\btheta)=q(\btheta)/c(\btheta)$, where $q(\btheta)$ is the unique unnormalized factorial Gaussian distribution with same moments as~(\ref{eq:moments}). The problem we address here is to approximate~$g(\btheta)$ when the moments are intractable.

\subsection{Laplace-style approximations}
\label{sec:laplace}

A strategy used in \cite{Smola-03,Yu-06} is to approximate~$g(\btheta)$ using the Laplace method applied to $c(\btheta)f(\btheta)$, {\em i.e.}, first find the value~$\btheta_\star$ that maximizes~$c(\btheta)f(\btheta)$, then approximate~$\log f(\btheta)=-\beta u(\btheta)$ by its second-order Taylor expansion at~$\btheta_\star$:
\begin{equation}
\label{eq:simple_laplace_approx}
\log g(\btheta) = -\beta 
\left[
u(\btheta_\star) 
+ \nabla u(\btheta_\star)^\top (\btheta-\btheta_\star)
+ \frac{1}{2} (\btheta-\btheta_\star)^\top 
{\rm diag}(\nabla \nabla^\top u(\btheta_\star))
(\btheta-\btheta_\star) 
\right]
,
\end{equation}
assuming that~$u(\btheta)$ is twice continuously differentiable. For a fully factorized approximation, only the diagonal elements of the Hessian of~$u(\btheta)$ need to be computed, hence the complexity of~(\ref{eq:simple_laplace_approx}) is linear in~$d$. 

However, the maximization step may be time consuming and we shall also consider a simpler variant, hereafter referred to as ``quick Laplace'', whereby the Taylor expansion is performed at the cavity center~$\bmu$ rather than at the maximizer~$\btheta_\star$ of~$c(\btheta)f(\btheta)$. Note that this makes the factor approximation independent from the cavity variance~$\S$.

\subsection{Gaussian quadrature}
\label{sec:gauss_quad}

As pointed out above, the cost derivatives may be discontinuous, or change abruptly in the cavity neighborhood, in which case they are not informative about the factor at the relevant scale of analysis. In such situations, the Laplace method can lead to poor factor approximations. An alternative is to use Gaussian quadrature to approximate the moments~(\ref{eq:moments}). For instance, \cite{Zoeter-05,Yu-06} use a precision-3 rule: 
\begin{equation}
\label{eq:quadrature_rule}
\hat{m}_i = 
\sum_{j=0}^{2d} w_j f(\btheta_j) \phi_i(\btheta_j)
\approx \int c(\btheta) f(\btheta) \phi_i(\btheta)d\btheta
,
\end{equation}
which involves a weighted average of $(2d+1)$ points:
$$
\left\{
\begin{array}{lll}
\displaystyle \btheta_0 = \bmu, & w_0 = 1 - \frac{d}{\gamma^2} & \\
\displaystyle \btheta_j = \bmu + \gamma \sigma_j \e_j, & w_j = \frac{1}{2\gamma^2}, & j=1,\ldots,d \\
\displaystyle \btheta_j = \bmu - \gamma \sigma_{j-d} \e_{j-d}, & w_j = \frac{1}{2\gamma^2}, & j=d+1,\ldots,2d,
\end{array}
\right
.
$$
where $\e_j$, for $j=1,\ldots,d$, denote the canonical vectors of $\mathbb{R}^d$, and $\gamma$ is a free parameter. As shown in \cite{Julier-00}, this quadrature rule is exact for polynomials up to degree~3, implying that it can recover the moments up to order~2 of a constant factor (since a factor which is both linear in~$\btheta$ and positive-valued on $\mathbb{R}^d$ has to be constant). 

The choice $\gamma=\sqrt{3}$ minimizes the error on the fourth-order moment of a standard Gaussian~\cite{Julier-00}, and the central weight $w_0$ is then negative for $d>3$. However, \cite{Yu-06} argue that non-negative weights should be used to guarantee non-negative second-order moments and adopt $\gamma = \sqrt{d}$. This leads to $w_0=0$, hence discarding the central sample point. I instead set $\gamma=\sqrt{d+0.5}$ to induce uniform weights and avoid zero weights. The quadrature rule~(\ref{eq:quadrature_rule}) then becomes formally similar to a Monte Carlo integral estimate using random points drawn independently from the cavity distribution~$c(\btheta)$, despite that the sampling points $\theta_j$, for $j=0,\ldots, 2d+1$, are chosen in a deterministic manner.

\subsection{Variational quadrature}
\label{sec:variational_quad}

A computational advantage of the precision-3 quadrature rule (\ref{eq:quadrature_rule}) is to be linear in~$d$, requiring only $(2d+1)$ evaluations of~$f(\btheta)$, but this may come at the price of fairly inaccurate moment estimates. The variational sampling method, which I originally proposed outside the context of~EP \cite{ijasp:13}, has the potential to improve over these estimates as it is exact for fully factorized Gaussian factors, just like Laplace-style approximations, yet relying on the same $(2d+1)$ evaluations as the precision-3 rule and not on differential calculus.

The key insight stems from recasting the moment-matching problem into that of minimizing the generalized Kullback-Leibler~(KL) divergence \cite{Minka-01,Minka-05}:
\begin{equation}
\label{eq:kl_div}
D( c f \| c g)
= 
\int 
c(\btheta)
\left[
f(\btheta) \log \frac{f(\btheta)}{g(\btheta)}
- f(\btheta)+g(\btheta)
\right] d\btheta
\end{equation}
over the set of unnormalized factorized Gaussian distributions, {\em i.e.}, functions of the form~$g(\btheta)=\exp[\balpha^\top \bphi(\btheta)]$ parameterized by~$\balpha \in \mathbb{R}^{2d+1}$, where~$\bphi(\btheta)$ is the vector-valued function with coordinate applications given by the monomials~(\ref{eq:monomials}). Dropping the terms independent from~$g$ in~(\ref{eq:kl_div}), we see that minimizing the KL~divergence with respect to~$\balpha$ is equivalent to minimizing the function:
\begin{eqnarray}
L(\balpha) & = &  
\int c(\btheta) 
\left[
-f(\btheta) \log g(\btheta) 
+ g(\btheta) 
\right] d\btheta
\nonumber\\
 & = & 
- \balpha^\top \int c(\btheta) f(\btheta) \bphi(\btheta) d\btheta
+ \int c(\btheta) e^{\balpha^\top \bphi(\btheta)} d\btheta
\label{eq:loc_cross_ent}
.
\end{eqnarray}

Therefore, instead of attempting at a direct moment evaluation~(\ref{eq:moments}), we may consider minimizing an empirical approximation to~(\ref{eq:loc_cross_ent}) based, for instance, on the precision-3 rule:
\begin{eqnarray}
\tilde{L}(\balpha) & = &  
\sum_{j=0}^{2d} 
w_j
\left[
-f(\btheta_j) \log g(\btheta_j)
+ g(\btheta_j)
\right]
\nonumber\\
 & = & 
- \balpha^\top \sum_{j=0}^{2d} 
w_j f(\btheta_j) \bphi(\btheta_j)
+ \sum_{j=0}^{2d} w_j e^{\balpha^\top \bphi(\btheta_j)}
\label{eq:surrogate_kl}
.
\end{eqnarray}

It turns out that (\ref{eq:surrogate_kl}) is a strictly convex function of~$\balpha$, see \cite{ijasp:13} for a proof (more generally, strict convexity requires that the number of sample points~$\btheta_j$ be at least as large as the number of real-valued moments, a condition satisfied in the present case since both are~$2d+1$). Therefore, $\tilde{L}(\balpha)$ has a unique minimizer~$\balpha_\star$, defining a factorized Gaussian distribution~$g_\star(\btheta)$ which is hopefully close to the actual moment-matching distribution and KL~minimizer. 

While $\balpha_\star$ lacks a closed-form expression, it can be tracked efficiently using Newton's method owing to the strict convexity of~$\tilde{L}(\balpha)$. The gradient and Hessian expressions are then needed:
\begin{eqnarray}
\nabla \tilde{L}(\balpha) & = & \sum_{j=0}^{2d} w_j \left[e^{\balpha^\top \bphi(\btheta_j)}-f(\btheta_j) \right]\bphi(\btheta_j), 
\label{eq:surrogate_gradient} \\
\nabla\nabla^\top \tilde{L}(\balpha) & = & \sum_{j=0}^{2d} w_j e^{\balpha^\top \bphi(\btheta_j)} \bphi_j(\btheta_j) \bphi_j(\btheta_j)^\top
. \nonumber 
\end{eqnarray}

It is important to realize that, in general, the optimal fit $g_\star(\btheta)$ does {\em not} match the moments approximated by a direct application of Gaussian quadrature~(\ref{eq:quadrature_rule}). In particular, if the factor~$f(\btheta)$ is a factorized Gaussian, then $g_\star = f$ and the moments are thus recovered {\em exactly} unlike using Gaussian quadrature. This can be seen from the gradient formula~(\ref{eq:surrogate_gradient}): if there exists $\balpha_0$ such that $f(\btheta)=\exp[\balpha_0^\top \bphi(\btheta)]$, then clearly $\nabla \tilde{L}(\balpha_0)=0$, meaning that $\balpha_\star=\balpha_0$. In some sense, variational quadrature bridges the gap between Laplace's method and Gaussian quadrature: it is error-free for factors belonging to the EP~approximating exponential family (like Laplace in the Gaussian case), and can handle non-continuously differentiable factors provided that the integrals~(\ref{eq:moments}) exist (like Gaussian quadrature). 

My implementation of Newton's method uses the Cholesky decomposition to invert Hessian matrices, hence the complexity of variational quadrature is in~$O(d^3)$. The computational overhead with respect to Gaussian quadrature therefore increases with~$d$, although it is partially compensated for by the fact that variational quadrature directly outputs an approximating factor~$g_\star(\btheta)$, while Gaussian quadrature requires dividing the moment-matching distribution by the cavity distribution, which becomes more costly as $d$~increases. Experimental comparisons of timing are provided next.

\section{Experiments}
\label{sec:experiments}

The different variants of FF-EP described in Section~\ref{sec:approx_schemes} were evaluated on real data from the UCI~repository (\url{https://archive.ics.uci.edu/ml/datasets.html}) to train linear binary classifiers according to different strategies: logistic regression, hinge regression, and a non-convex approximation to the minimization of the mean classification error.

\subsection{Implementation}

The FF-EP algorithm and the numerical approximation schemes described in Section~\ref{sec:approx_schemes}, respectively referred to as Laplace method (LA), quick Laplace method (QLA), Gaussian quadrature (GQ) and variational quadrature (VQ) in this section, were implemented in scientific Python (\url{www.scipy.org}). Newton's method was used for both the optimization step in~LA (see Section~\ref{sec:laplace}) and for VQ (see Section~\ref{sec:variational_quad}) with a tolerance of~$10^{-5}$ on relative parameter variations.


Each UCI~dataset was formatted as a list of examples of the form $(y_k, \x_k)$, for $k=1,2,\ldots,N$, where $y_k \in \{-1,1\}$ was a binary label and $\x_k\in \mathbb{R}^d$ a feature vector. A constant value (baseline) was appended to all feature vectors. In a pre-processing step, all features were normalized to unit Euclidean norm after subtracting the mean across examples, except for the baseline.

In all experiments, the FF-EP algorithm incorporated a fixed Gaussian factor with zero mean and scalar variance matrix $25 {\bf I}_d$ modeling a prior distribution on the linear classification parameter vector~$\btheta\in \mathbb{R}^d$. Factors were defined according to the chosen training cost (see below) by summing up the costs over mini-batches and exponentiating the result as in~(\ref{eq:boltzmann_dist}) using a fixed ``inverse temperature'' parameter~$\beta=1$. The corresponding factor approximations in FF-EP were initialized as identically equal to one ({\em i.e.}, with infinite variances).

\subsection{Classification cost functions}

Different cost functions were considered for linear classification training:

\paragraph{Logistic loss.} This choice leads to a smooth function of~$\btheta$,
$$
u_k(\btheta) = \log (1 + e^{-y_k \btheta^\top \x_k} ).
$$

\paragraph{Hinge loss.} This is the loss classically used to train support vector machines \cite{Ertekin-11}, 
$$
u_k(\btheta) = \max(0, 1 -y_k \btheta^\top \x_k ).
$$

\begin{figure}[!ht]
\begin{center}
\includegraphics[width=.45\textwidth]{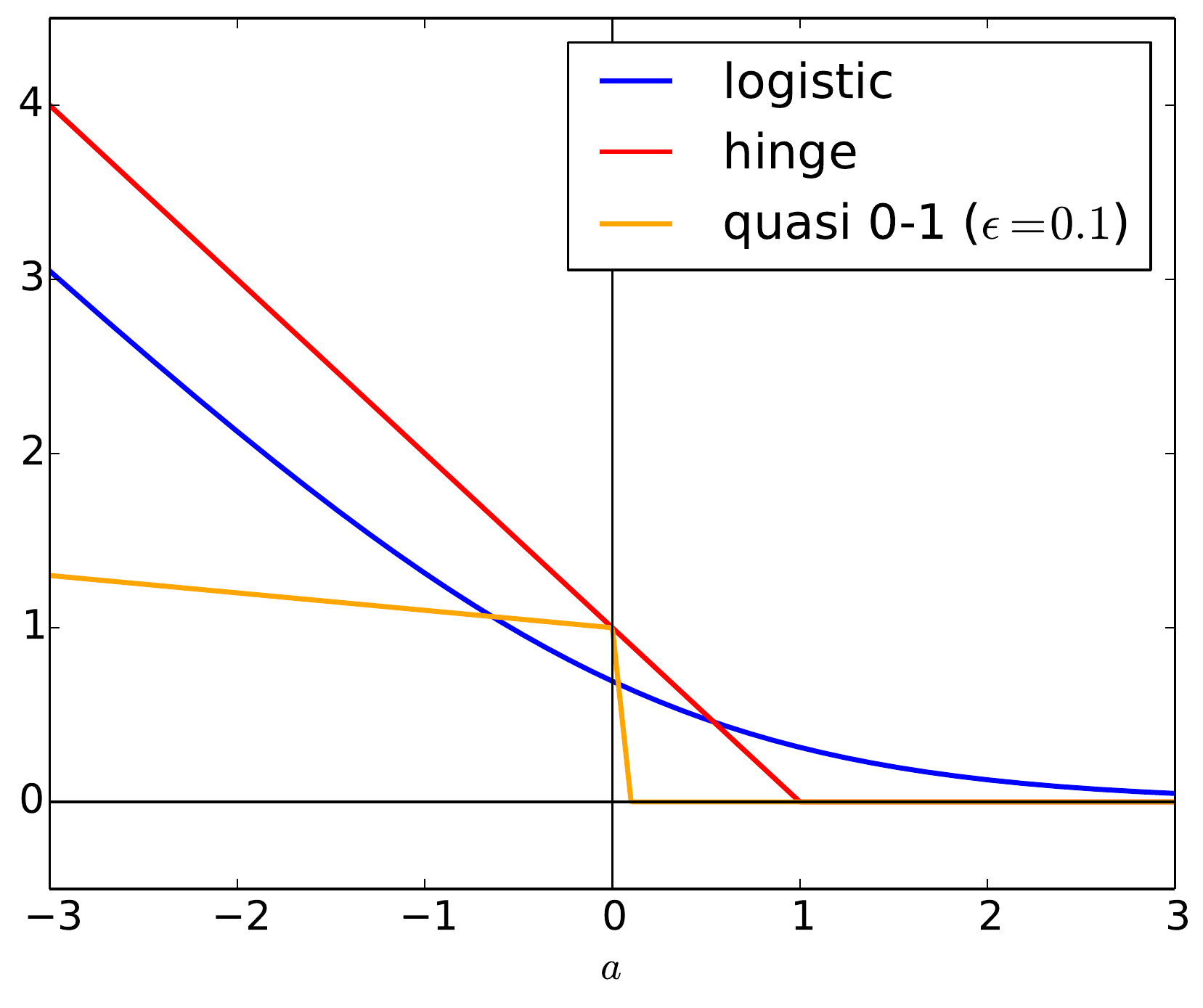}
\end{center}
\caption{Loss functions considered for binary linear classification training.}
\label{fig:loss_funcs}
\end{figure}

\paragraph{Quasi 0--1 loss.} In some applications, the desired training objective is to minimize the mean classification error using the 0--1 loss. However, because the 0--1 loss is discontinuous and turns out to make the proposed FF-EP variants unstable, it is approximated by a continuous non-convex function:
$$
u_k(\btheta) = \ell_\epsilon(y_k \btheta^\top \x_k),
\qquad
{\rm with}
\quad 
\ell_\epsilon(a) = 
\left\{
\begin{array}{lll}
1 - \epsilon a & {\rm if} & a < 0 \\
1 - a/\epsilon & {\rm if} & 0 \leq a < \epsilon \\
0 & {\rm if} & a > \epsilon 
\end{array}
\right.
,
$$
for some constant~$\epsilon>0$. While the case~$\epsilon=1$ corresponds to the hinge loss, $\ell_\epsilon(a)$ provides a tighter approximation to the 0--1 loss for smaller~$\epsilon$, see Figure~\ref{fig:loss_funcs}. In these experiments, I set $\epsilon=0.1$. Note that the quasi 0--1 loss differs from the ``ramp loss'' \cite{Ertekin-11}.

Note that both the hinge loss and the quasi 0--1 loss derivatives are singular, respectively on the hyperplane $\{\btheta^\top \x_k = y_k\}$ and on the hyperplanes $\{\btheta^\top \x_k = 0\}$ and $\{\btheta^\top \x_k = \epsilon y_k\}$. Derivatives at discontinuities were conventionally defined by extending $\ell'_\epsilon(a)$ at $a=0$ and $a=\epsilon$ by the half sum of its left and right limits.

\subsection{Small datasets}

I considered three UCI~classification datasets containing less than $N=1000$ examples (Haberman's survival, Ionosphere, Wisconsin diagnostic breast cancer). For the different training objectives, the question is whether the numerically approximated FF-EP algorithm converges (knowing that convergence is not even warranted in the case of explicit updates \cite{Minka-05}) and, if so, how fast and how close to the global cost minimizer.

Mini-batches of size~$s=10$ were used in these experiments, which enabled to loop several times over the whole set of mini-batches at reasonable computational cost. Using relatively small mini-batches, FF-EP searches for a consensus between many ``weak learners'' while giving several opportunities to each learner to revise its knowledge during the process.

\begin{figure}[!ht]
\begin{center}
\includegraphics[width=.32\textwidth]{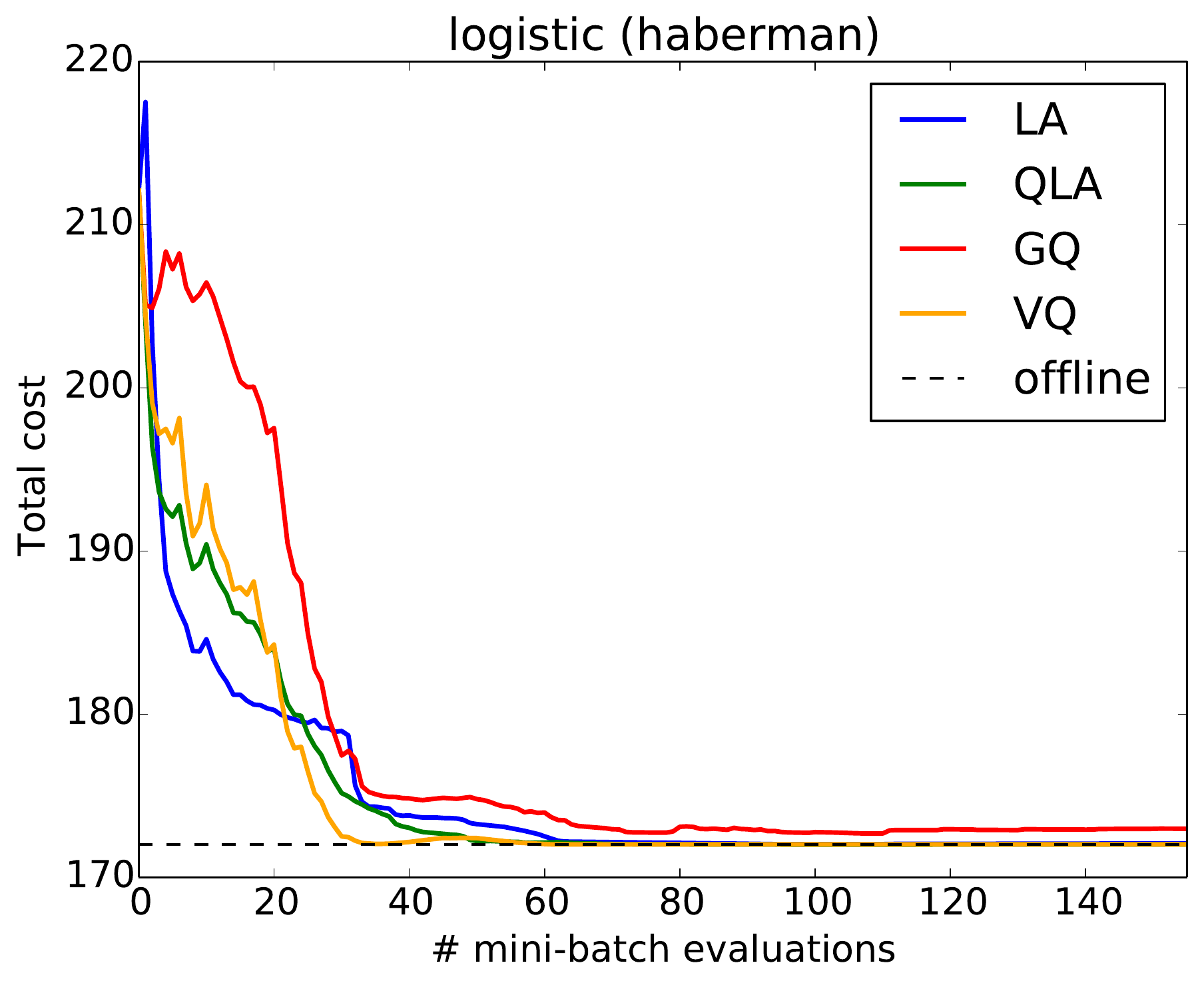}
\includegraphics[width=.32\textwidth]{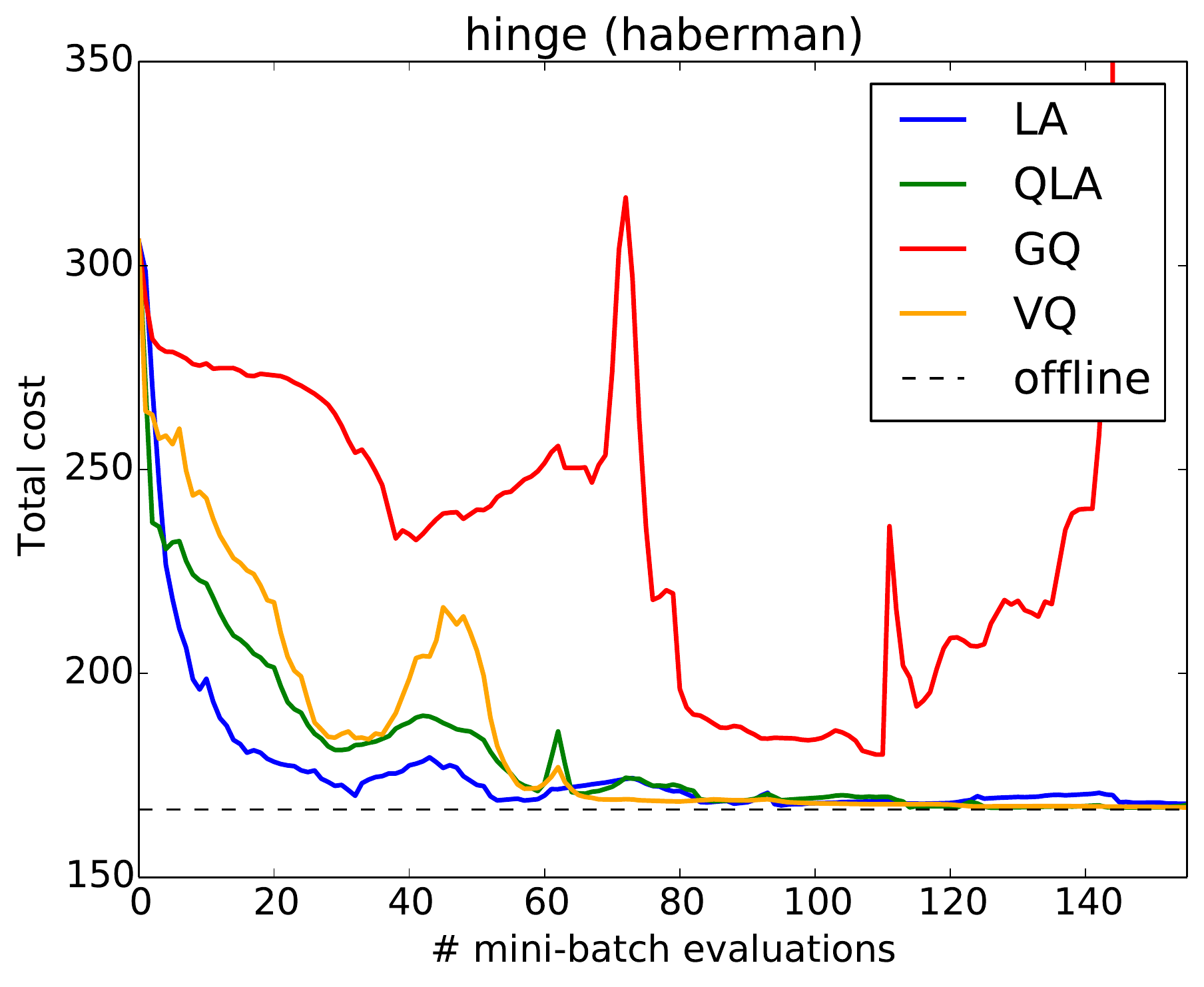}
\includegraphics[width=.32\textwidth]{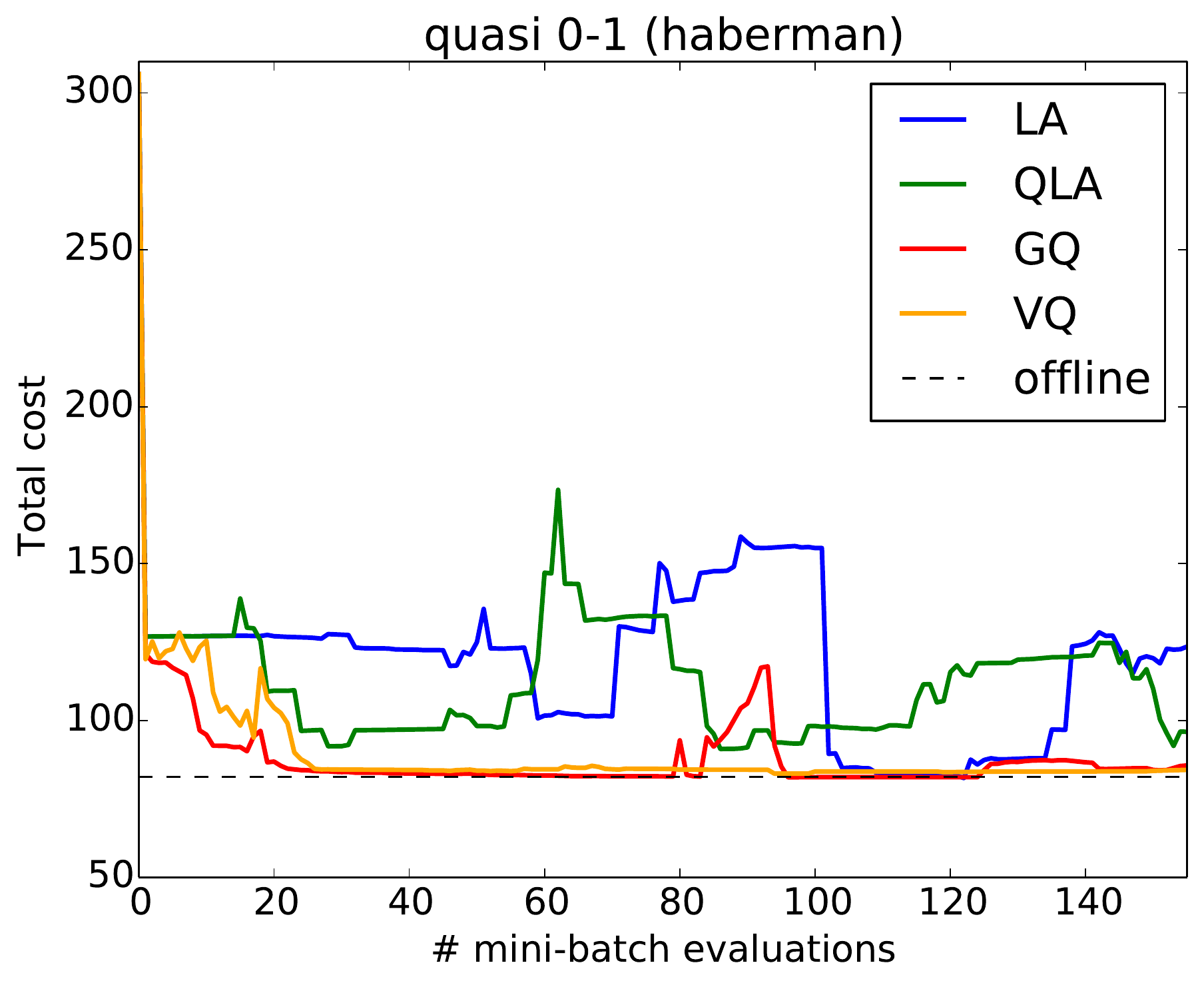}
\includegraphics[width=.32\textwidth]{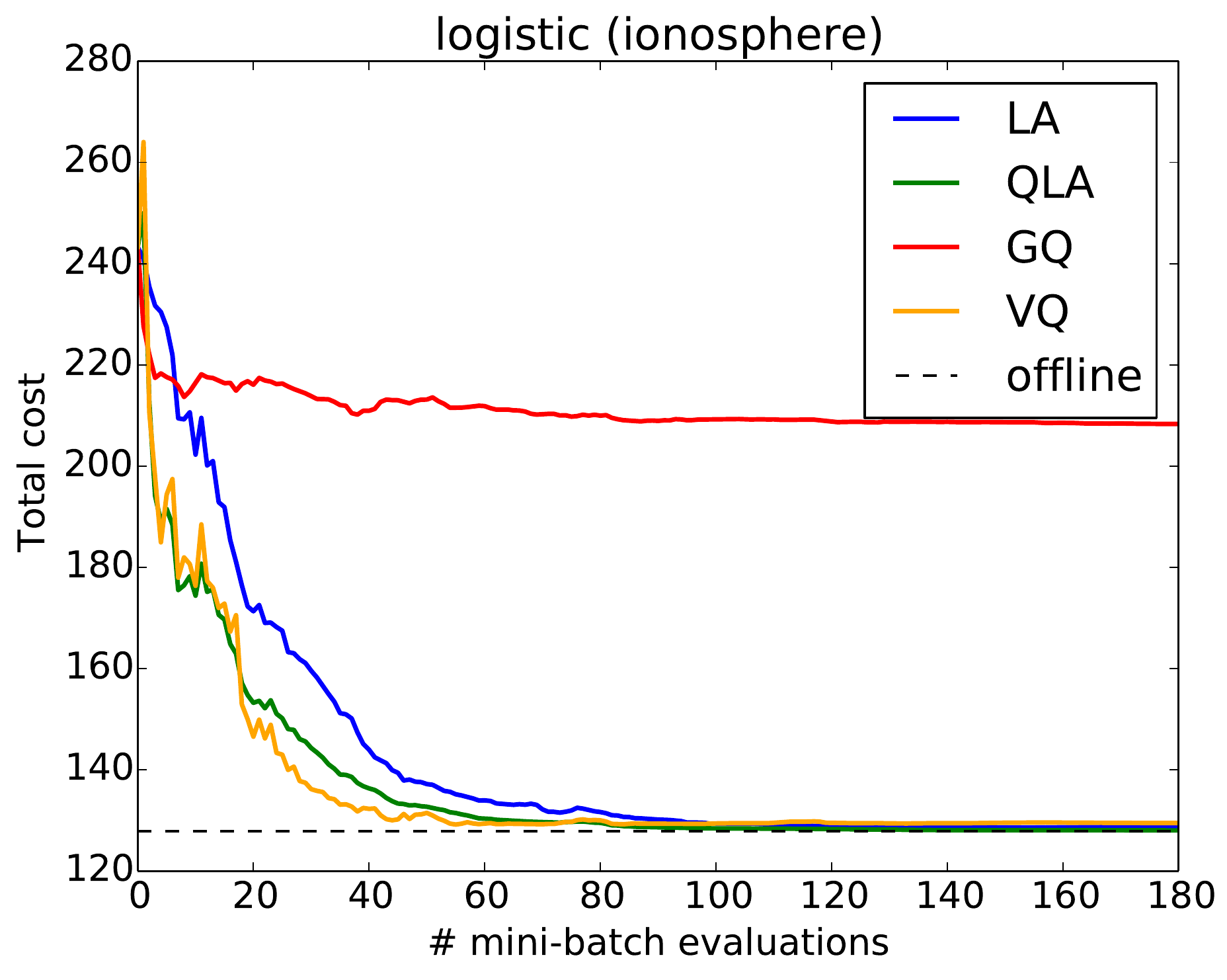}
\includegraphics[width=.32\textwidth]{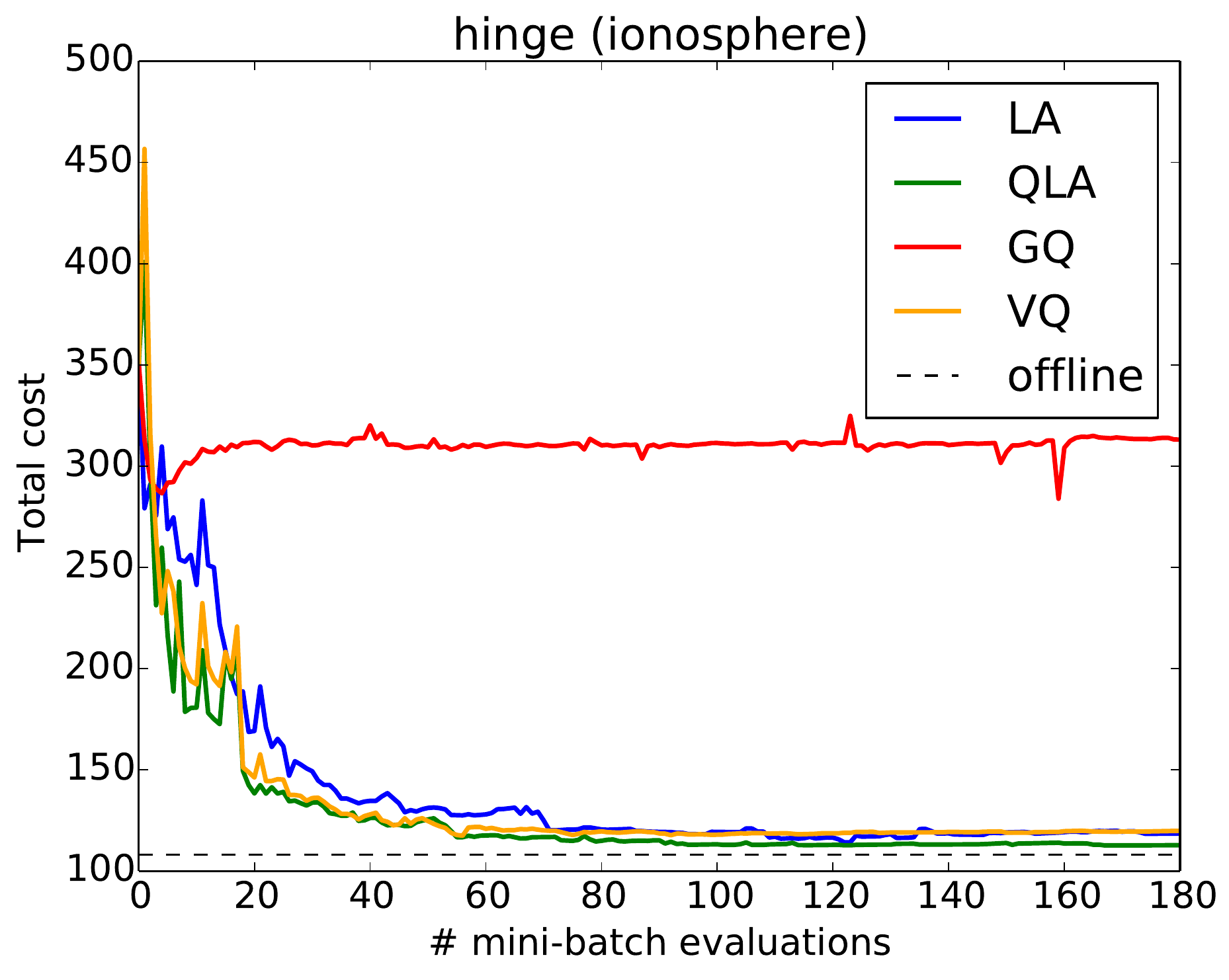}
\includegraphics[width=.32\textwidth]{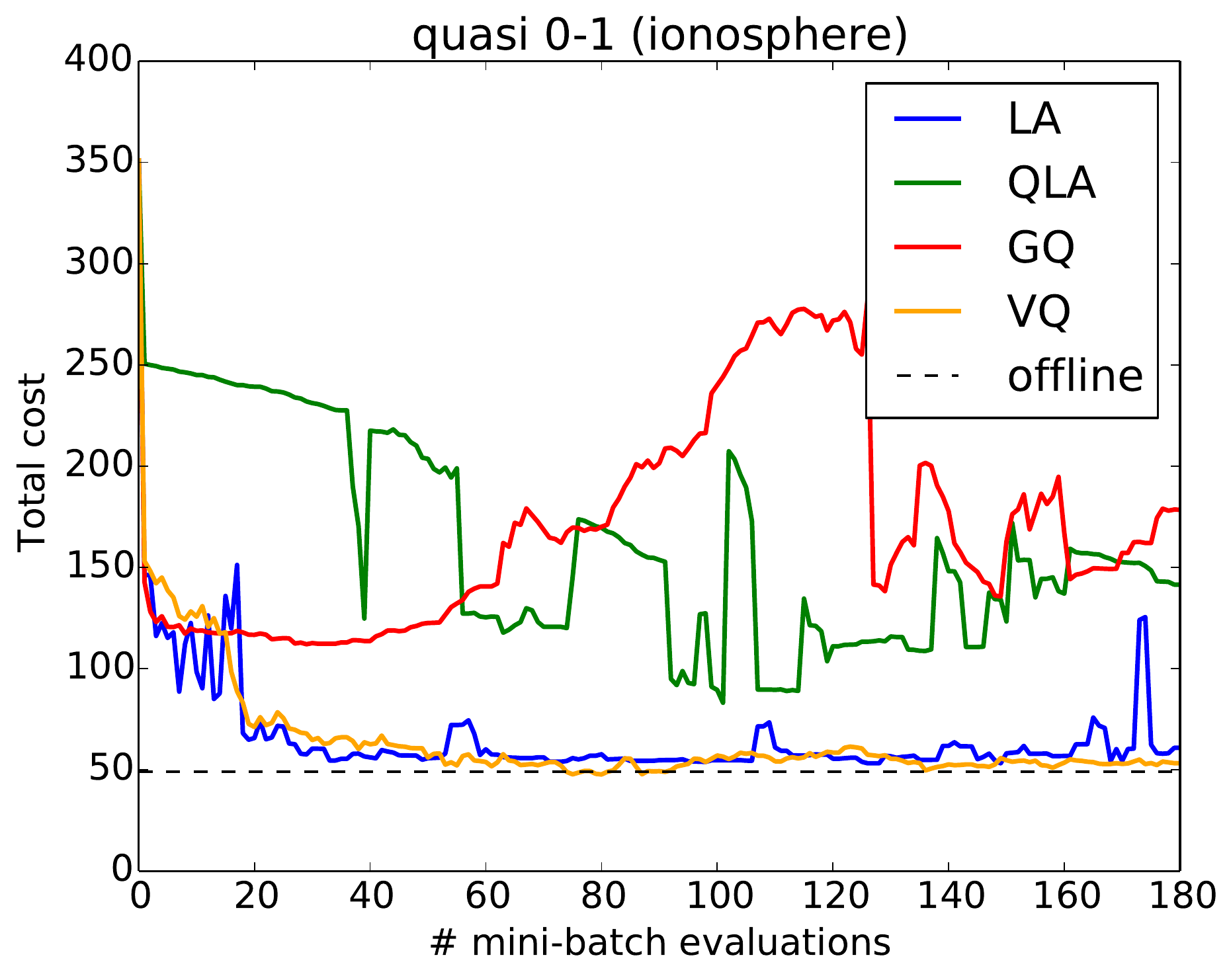}
\includegraphics[width=.32\textwidth]{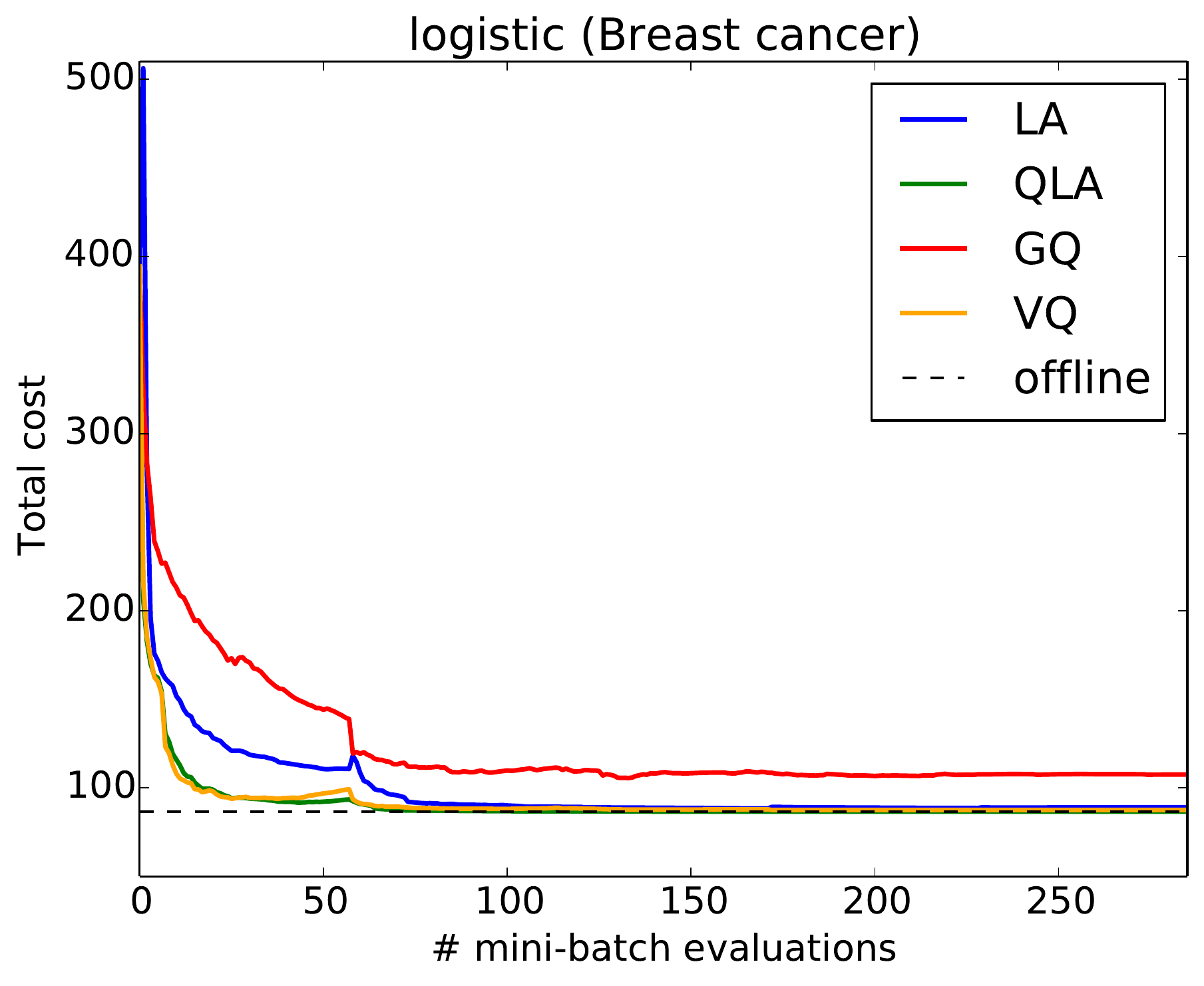}
\includegraphics[width=.32\textwidth]{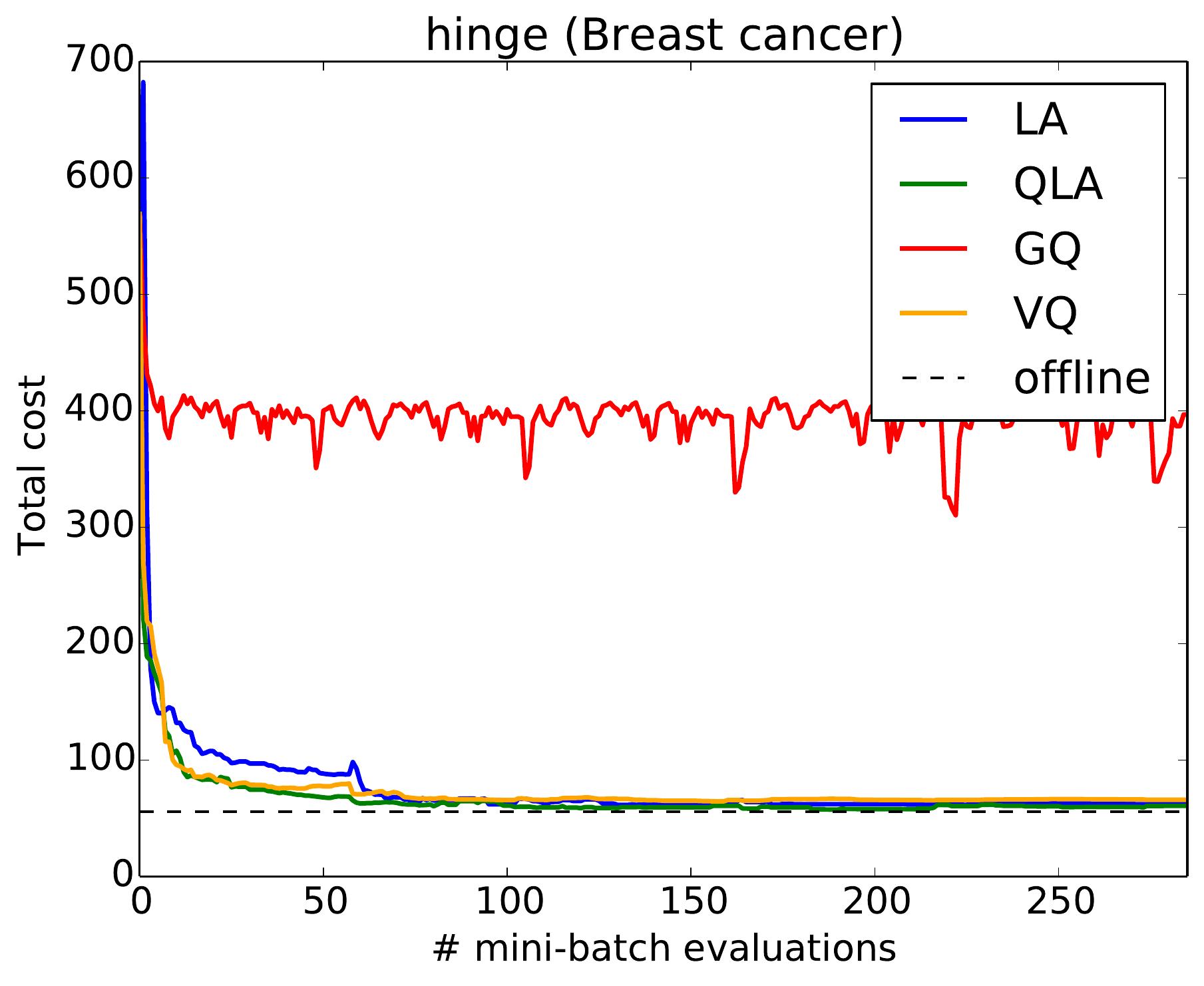}
\includegraphics[width=.32\textwidth]{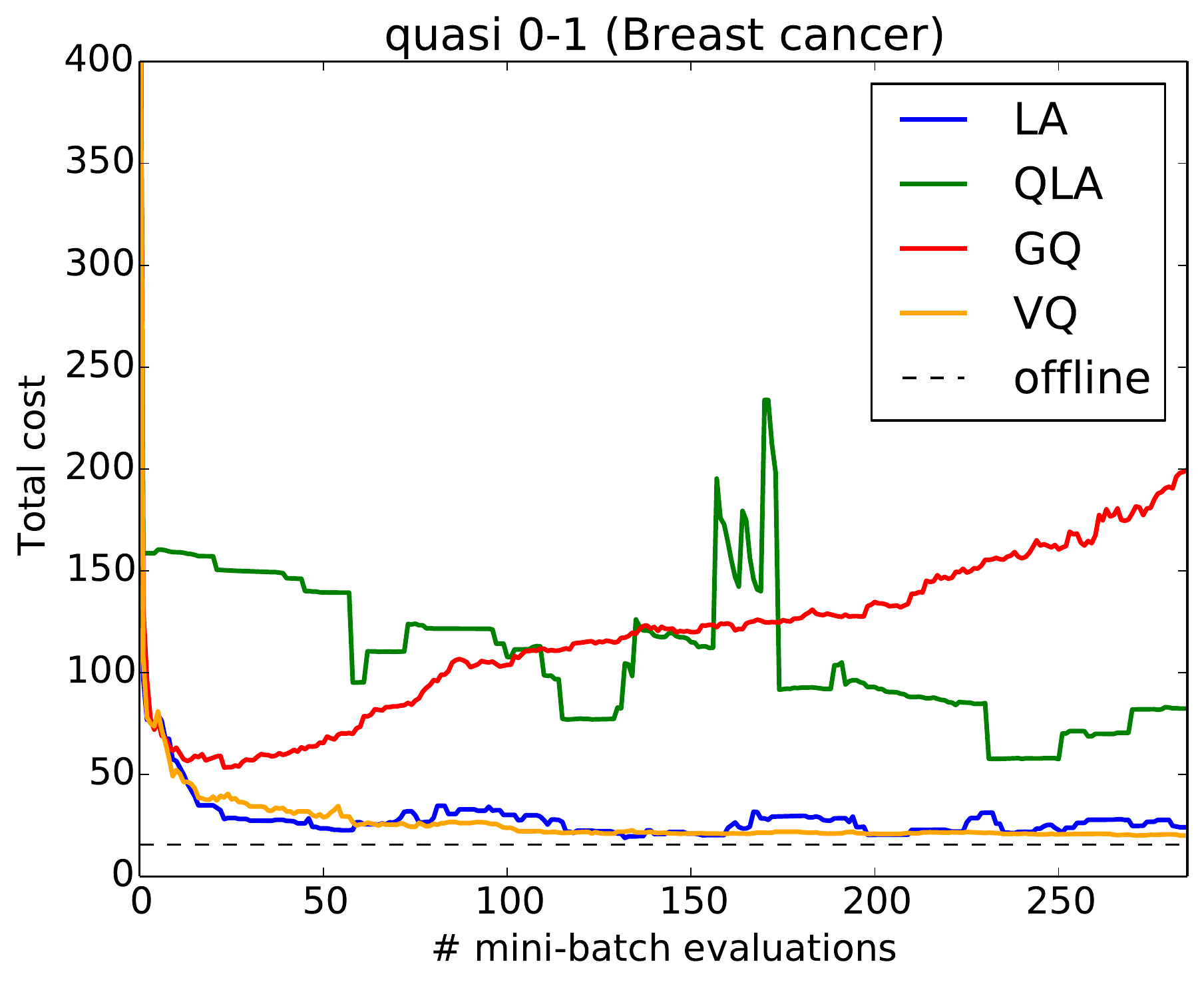}
\end{center}
\caption{Evolution of total classification cost over FF-EP~updates for small datasets (from top to bottom: Haberman, ionosphere, breast cancer), different loss functions (from left to right: logistic, hinge and quasi 0--1), and different approximation schemes.}
\label{fig:small_datasets}
\end{figure}

The curves in Figure~\ref{fig:small_datasets} show the variations of the total cost (computed offline) across mini-batch updates, for the different datasets, cost functions, and FF-EP approximation schemes. For comparison, the regression parameters~$\btheta$ minimizing the respective costs were also computed offline using Newton's method for the logistic loss, and Powell's method (initialized from the logistic solution) for both the hinge and quasi~0--1 losses. 

Five loops proved to be more than enough to achieve convergence in most cases using FF-EP with VQ, which was clearly the most stable among the tested approximation methods in all scenarios, and converged to close-to-optimal parameters. The slight residual fluctuations observed for the ionosphere dataset  could be due to the combined effect of non-convexity and a relatively small number~$N$ of examples compared to the number~$d$ of features in this dataset. Both Laplace-style approximations (LA and QLA) converged well for logistic and hinge regression, although the hinge loss is not continuously differentiable, but were both unstable in quasi~0--1 regression. GQ showed oscillatory behavior or local convergence in all cases, except to some extent for logistic regression in small dimension (Haberman dataset, $d=4$).

While QLA performed comparably with LA, it was much faster, as shown by the timings reported in Table~\ref{tab:small_datasets}. This is due to the function evaluations involved in the optimization step of~LA, and suggests that such optimization may not be needed in~EP. VQ entailed some computational overhead (depending on the parameter dimension~$d$) compared to both QLA and GQ, but was also significantly faster than LA.

\begin{table}[!h]
  \begin{center}
    \begin{tabular}{|c|c|c|c|c|c|c|c|c|}
      \hline
      \multirow{2}{*}{Dataset} & \multirow{2}{*}{$N$} & \multirow{2}{*}{$d$} & \multirow{2}{*}{$s$} & \multicolumn{5}{c|}{Mean time per mini-batch (ms)} \\
      & & & & Model & LA & QLA & GQ & VQ \\  
      \hline\hline
      \multirow{3}{*}{Haberman} & 
      \multirow{3}{*}{306} & \multirow{3}{*}{4} & \multirow{3}{*}{10} & 
      Logistic & 4.78 & 0.59 & 0.34 & 1.05 \\
      & & & & Hinge & 5.61 & 0.59 & 0.32 & 1.03 \\
      & & & & Quasi 0--1 & 11.86 & 0.85 & 0.41 & 1.68\\
      \hline
      \multirow{3}{*}{Ionosphere} & 
      \multirow{3}{*}{351} & \multirow{3}{*}{33} & \multirow{3}{*}{10} & 
      Logistic & 8.11 & 0.63 & 0.44 & 2.22 \\
      & & & & Hinge & 7.67 & 0.67 & 0.40 & 5.56\\
      & & & & Quasi 0--1 & 17.15 & 0.91 & 0.50 & 6.38\\
      \hline
      \multirow{3}{*}{Breast cancer} & 
      \multirow{3}{*}{569} & \multirow{3}{*}{31} & \multirow{3}{*}{10} & 
      Logistic & 9.86 & 0.63 & 0.40 & 1.59 \\
      & & & & Hinge & 7.86 & 0.65 & 0.38 & 1.82\\
      & & & & Quasi 0--1 & 12.82 & 0.91 & 0.49 & 1.81\\
      \hline
      \multirow{3}{*}{Bank marketing} & 
      \multirow{3}{*}{45211} & \multirow{3}{*}{43} & \multirow{3}{*}{100} & 
      Logistic & 95.33 & 9.85 & 4.26 & 8.26 \\
      & & & & Hinge & 117.39 & 10.37 & 3.91 & 9.64\\
      & & & & Quasi 0--1 & 323.56 & 15.62 & 5.78 & 10.36 \\
      \hline
      \multirow{3}{*}{Census income} & 
      \multirow{3}{*}{32561} & \multirow{3}{*}{101} & \multirow{3}{*}{100} & 
      Logistic & 191.53 & 12.20 & 8.14 & 33.97 \\
      & & & & Hinge & 230.76 & 12.49 & 7.20 & 37.98\\
      & & & & Quasi 0--1 & 628.13 & 17.87 & 9.10 & 33.93 \\
      \hline
    \end{tabular}
   \end{center}
  \caption{Dataset characteristics and timing on a standard single processor of different FF-EP versions for binary classification experiments. }
  \label{tab:small_datasets}
\end{table}

\subsection{Large datasets}

Both the Bank Marketing and Census income UCI~datasets were also considered for the empirical evaluation of numerical FF-EP approximations. For such training datasets containing several thousands of examples, we can afford larger mini-batches while keeping a sufficiently large number of mini-batches to take advantage of information redundancy in the data. This is motivated by the fact that larger mini-batches are presumably stronger learners. 

In both datasets, we chose mini-batches of size~$s=100$. Since there are about 100~times more examples than in the ``small datasets'', both the size and number of mini-batches were roughly 10~times larger than in the previous classification experiments. A single loop over mini-batches was performed. In this case, memory load can be substantially reduced as there is no need to keep the approximate factors in memory; since they are initialized as uniform distributions, the cavity distribution can be updated on-the-fly as each mini-batch is visited only once. 

\begin{figure}[!ht]
\begin{center}
\includegraphics[width=.32\textwidth]{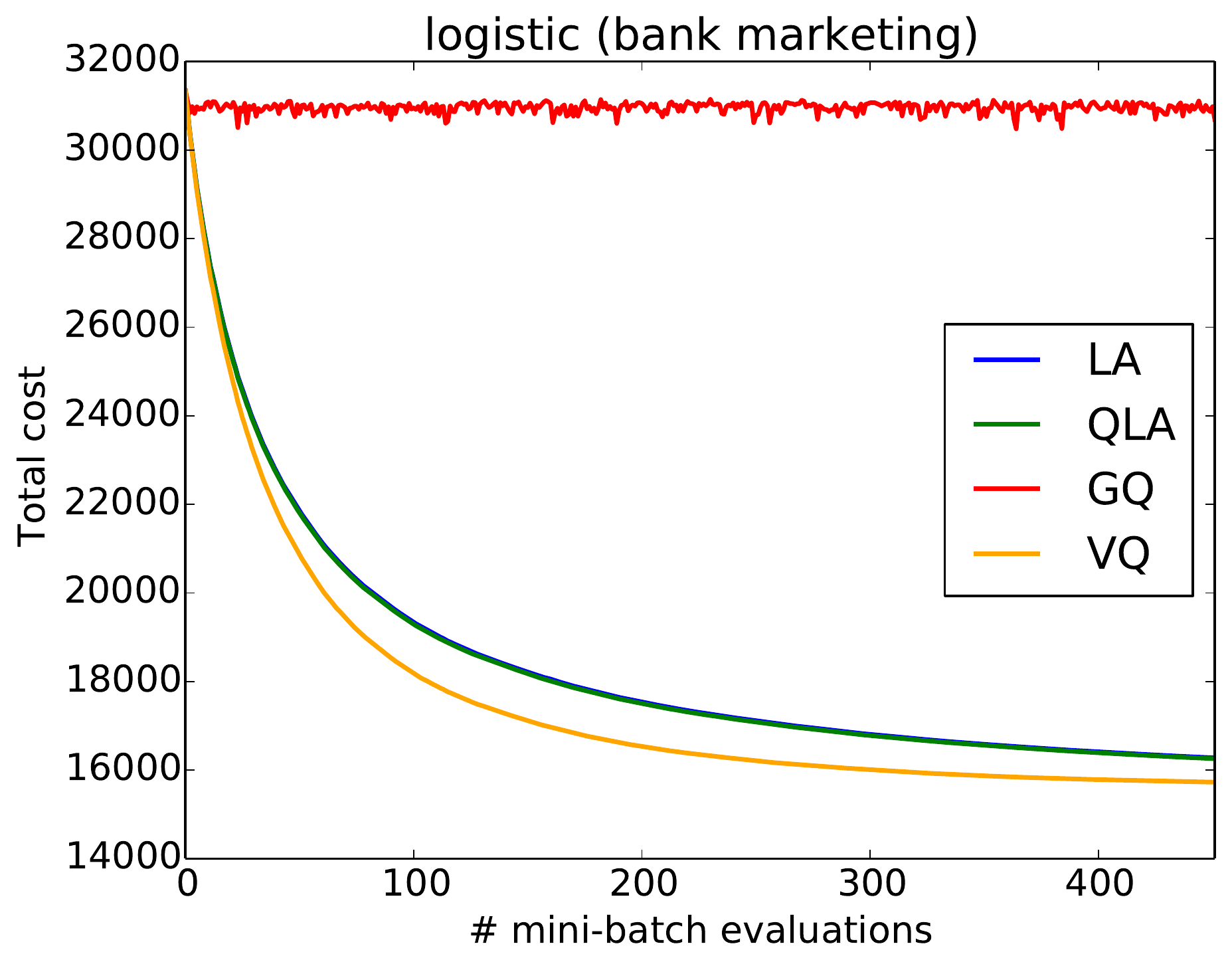}
\includegraphics[width=.32\textwidth]{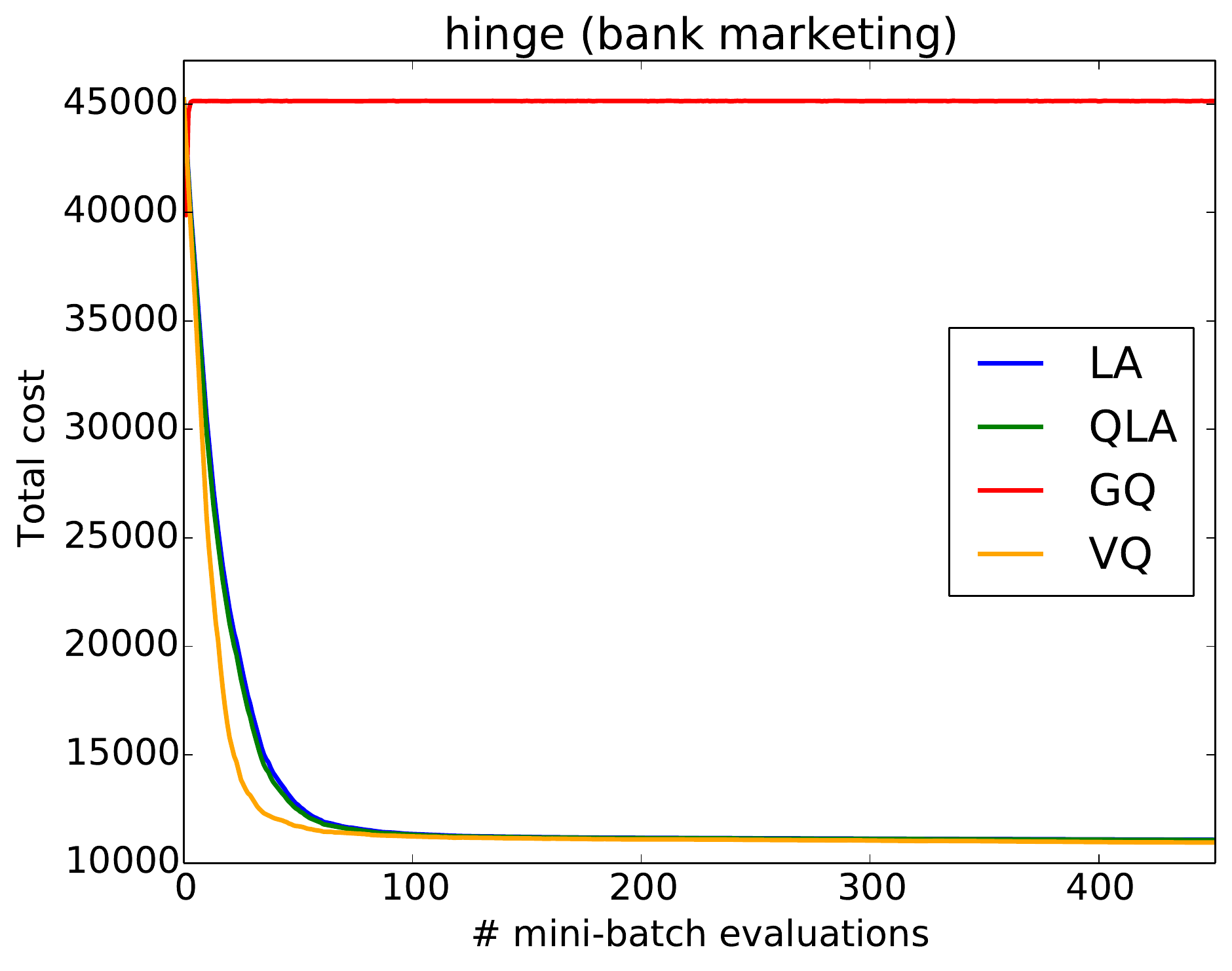}
\includegraphics[width=.32\textwidth]{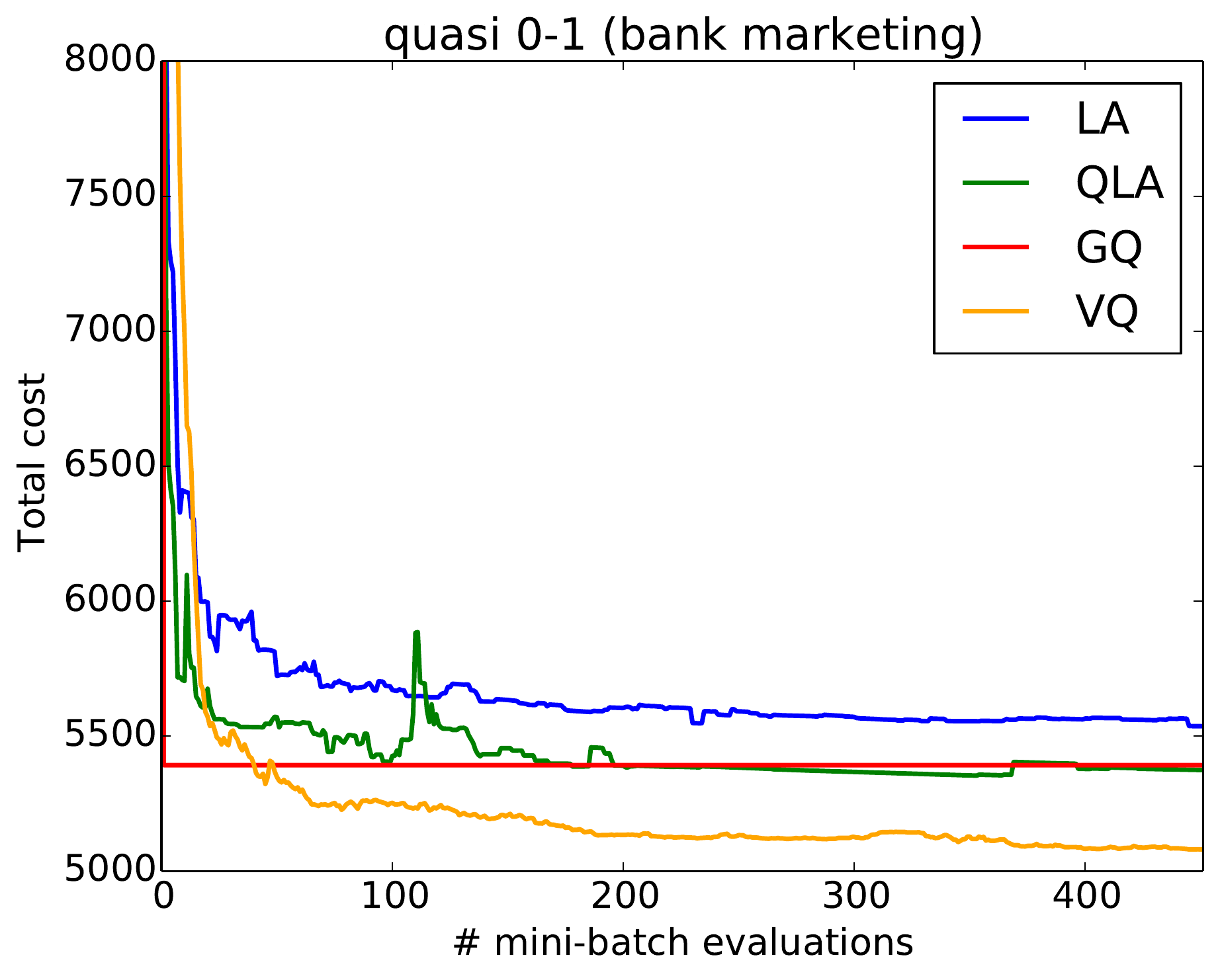}
\includegraphics[width=.32\textwidth]{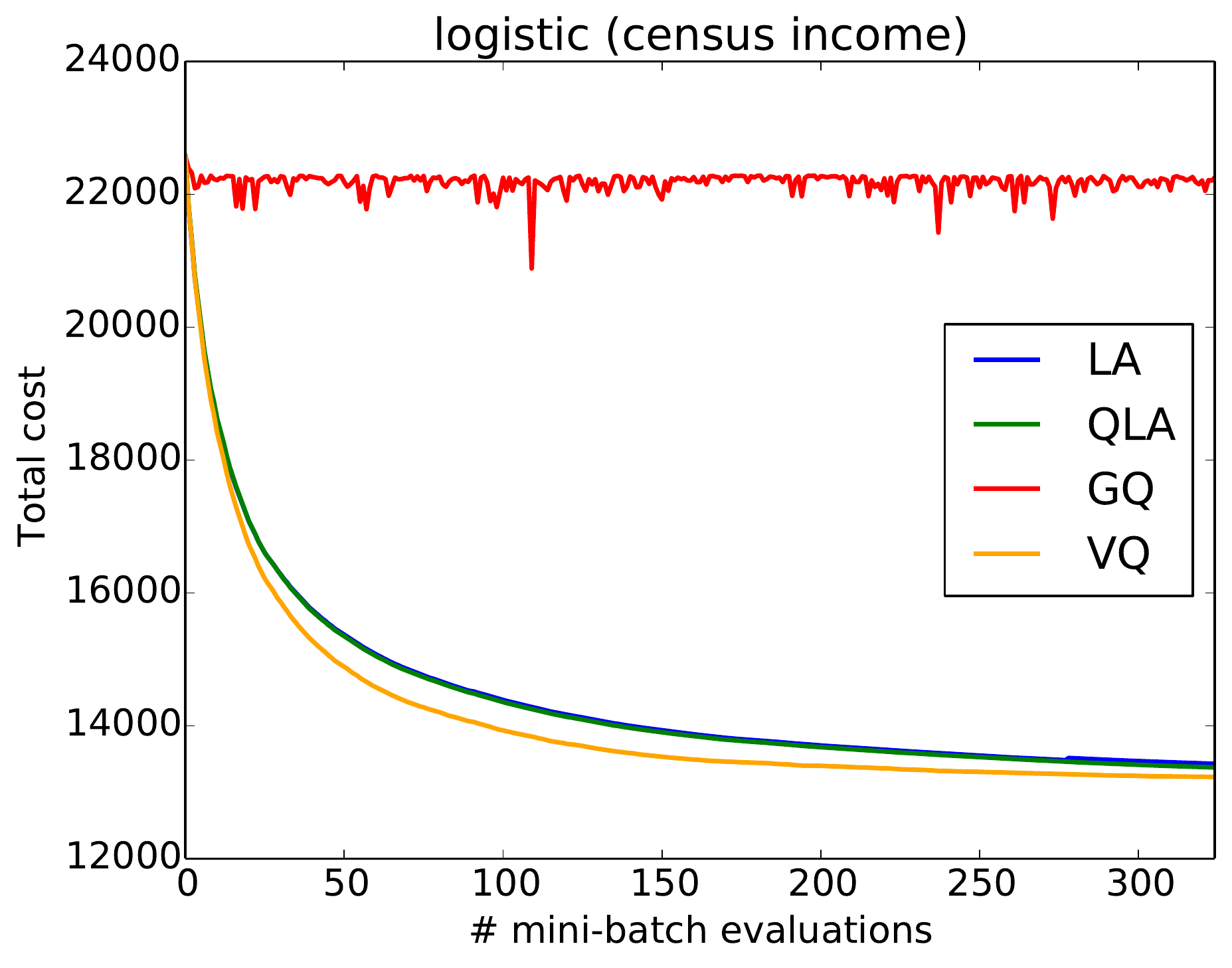}
\includegraphics[width=.32\textwidth]{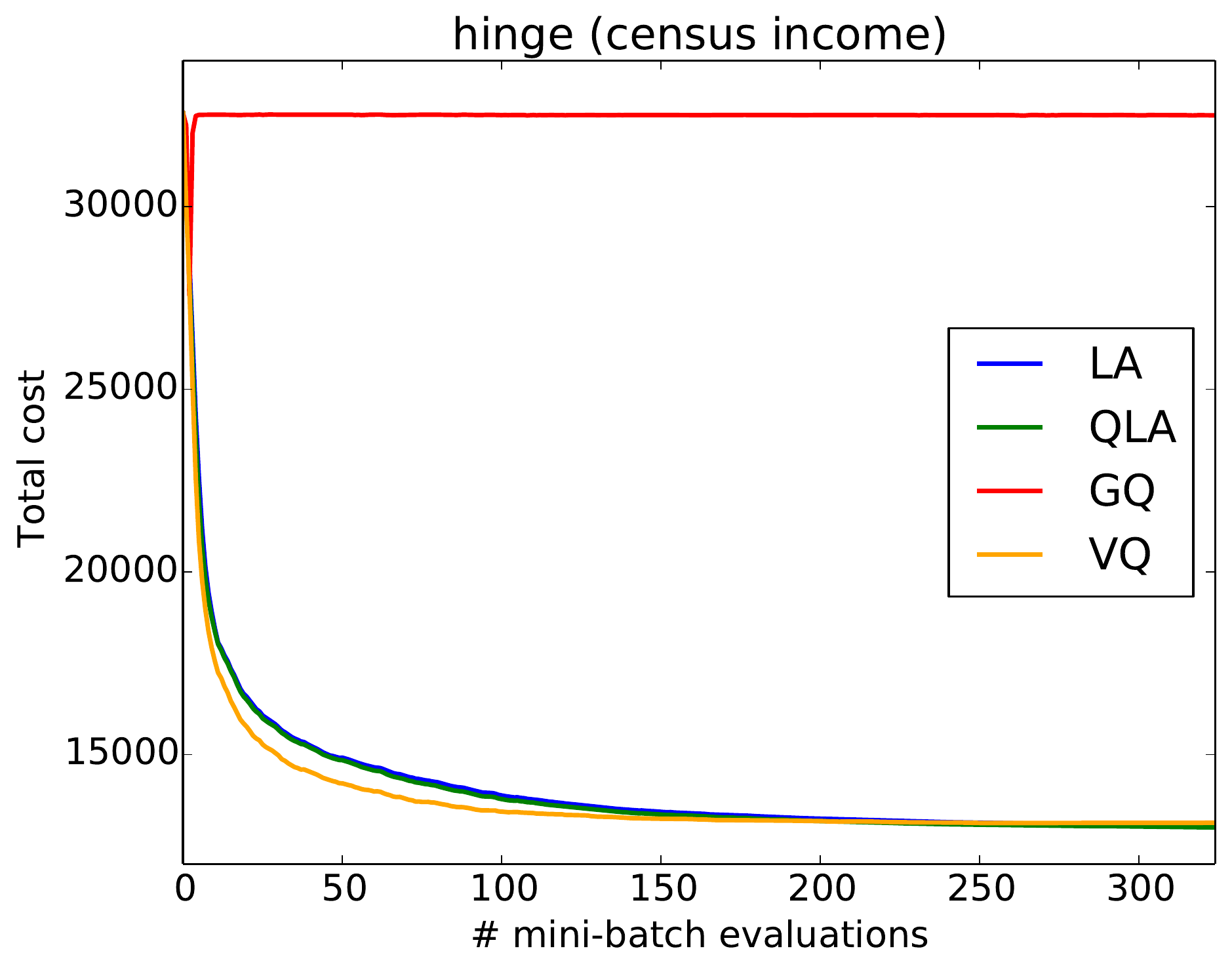}
\includegraphics[width=.32\textwidth]{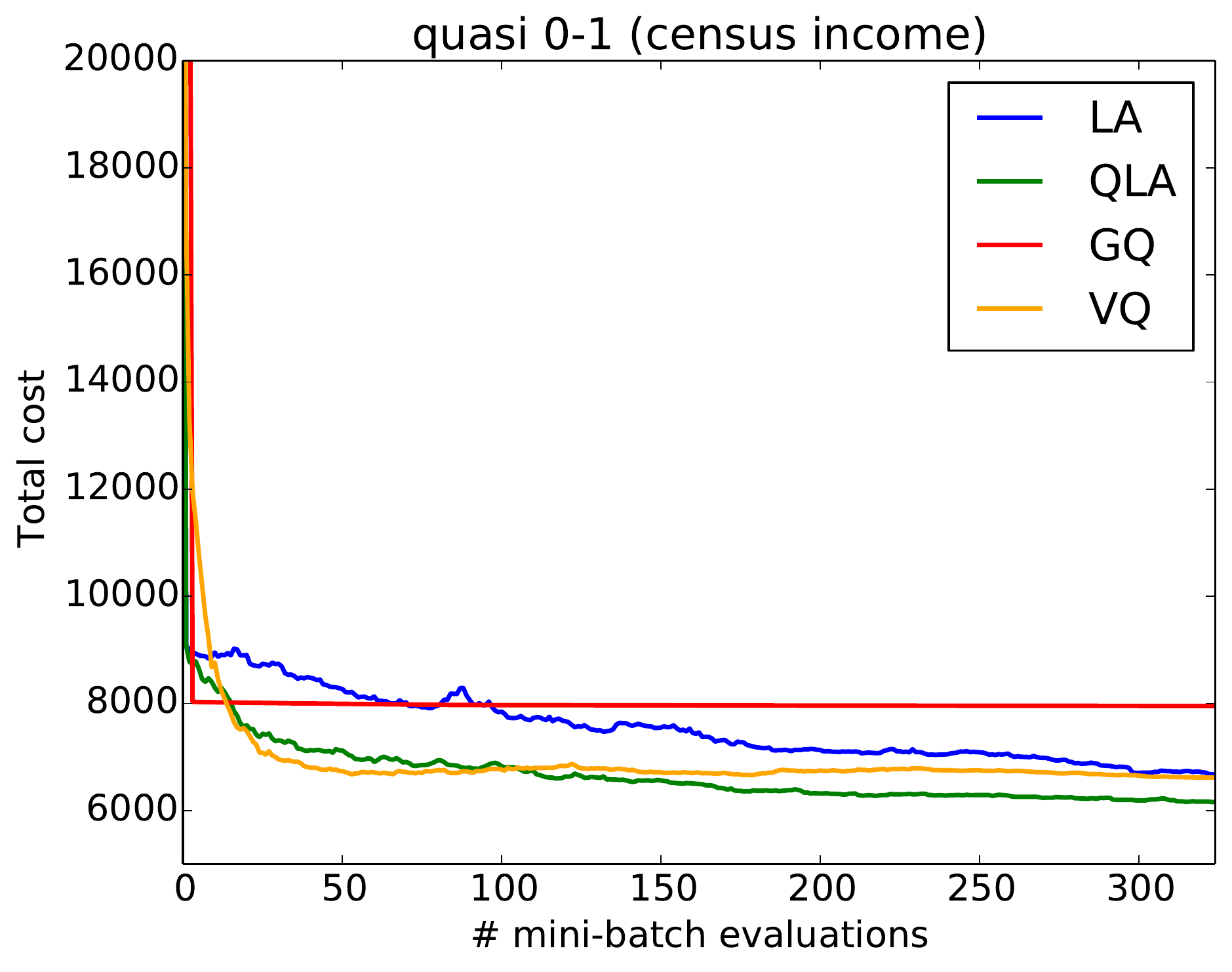}
\end{center}
\caption{Evolution of total classification cost over FF-EP~updates for large datasets (top: Bank marketing, bottom: Census income), different loss functions (from left to right: logistic, hinge and quasi 0--1), and different approximation schemes.}
\label{fig:large_datasets}
\end{figure}

Evolution curves of the same type as for the small dataset experiments are shown in Figure~\ref{fig:large_datasets}. In these experiments, the offline minimizations were not computed due to prohibitive computational cost. Both Laplace-style methods proved more stable and more similar to VQ in quasi 0--1 regression, while GQ had a strong tendency to underestimate factor variances and quickly get stuck on a sub-optimal solution. 

The better behavior of Laplace-based FF-EP can be explained by the larger number of factors (mini-batches) in these experiments, which helped stabilizing FF-EP updates, along with the fact that factors based on larger mini-batches tend to have more regular shapes. In the cases of logistic and hinge regression, convergence using either LA, QLA, or VQ occurred much before all mini-batches were visited. For the quasi 0--1 loss, QLA achieved a slightly lower cost than VQ, but did not converge as smoothly as~VQ.

Table~\ref{tab:small_datasets} shows that VQ ran slightly faster than QLA for the bank marketing data ($d=43$), and about 2--3~times slower for the  census income data ($d=101$), consistently with theory since the complexities of VQ and QLA are respectively in $O(d^3)$ and $O(d)$. LA was, again, very slow without showing better performance than QLA and VQ.

\section{Discussion}

We discussed and tested several methods to approximate EP~updates when analytically intractable: Laplace-style approximations, standard Gaussian quadrature and deterministic variational sampling, or variational quadrature. All  methods are widely applicable as they only require that the factors can be evaluated. Laplace-style methods further require evaluating the first- and second-order derivatives: using analytic expressions is generally preferred, but derivatives can also be approximated using finite differences, yielding the same complexity as the precision-3 quadrature rule considered here.

Experimental results confirm that the Laplace-based approximation works better for EP than low-order Gaussian quadrature, as previously reported by others \cite{Yu-06}, yet it is also significantly slower. Much computation time can be saved by avoiding the optimization step in the Laplace method, which does not seem to hamper performance in practice. However, both Laplace-based FF-EP algorithms were found to be sometimes unstable with non-smooth distributions. While the Laplace method is local in essence, both Gaussian quadrature and variational quadrature perform factor approximations at a coarser scale adaptively determined by the EP~algorithm via the cavity distribution. This makes them, in principle, more robust to sudden changes in factor slope or curvature, which may confuse Laplace-style methods while being irrelevant to the overall function approximation problem.

The precision-3 Gaussian quadrature rule proved, however, too inaccurate to enable good convergence of the FF-EP algorithm in practice. While this problem could potentially be overcome using higher order quadrature rules or Monte Carlo methods at the price of increased computation time, the refinement provided by variational quadrature without sampling additional points led to a dramatic improvement. The intuitive reason is that variational quadrature is designed to be exact for Gaussian factors, and is thus generally more accurate than Gaussian quadrature, which is only exact for constant factors (see Section~\ref{sec:gauss_quad}).

One drawback of variational quadrature is to scale unfavorably with parameter dimension. Using Newton's method in the fitting step, the complexity is in $O(d^3)$, while both quick Laplace and Gaussian quadrature are linear in~$d$. A possibility to reduce the complexity to linear would be to use an approximate minimization strategy. However, in moderate parameter dimension ($d<100$), the tested implementation proved experimentally faster than the Laplace method, which seems to be the current standard for numerical~EP \cite{Smola-03,Yu-06}, and showed an order of computation time comparable with both quick Laplace and Gaussian quadrature, two methods chosen for computational speed. 

The practical recommendation suggested by this work is to use the quick Laplace method in applications where the factors vary smoothly in~$\btheta$ and speed is a requirement; otherwise, variational quadrature seems to be preferable. Also note that increasing the size of mini-batches is a way to reduce the computational overhead of variational quadrature with respect to quick Laplace.


\end{document}